\documentclass[USenglish,twocolumn]{article}
\usepackage[utf8]{inputenc}
\usepackage[big,online]{dgruyter}
\usepackage[sort&compress,square,numbers]{natbib}
\usepackage{times}

\usepackage[separate-uncertainty  = true,
uncertainty-separator =  {\,},
output-decimal-marker = {.},
multi-part-units = single,
range-units = brackets,
range-phrase = {\,--\,}]{siunitx}

\usepackage{subcaption}
\usepackage{svg}
\usepackage{graphicx}
\usepackage{xcolor}

\usepackage{siunitx}

\usepackage{environ}

\usepackage{xargs}

\begin{document}

  \pagenumbering{gobble}

\title{A Calibration Approach for Elasticity Estimation with Medical Tools}

\runningtitle{Calibration of Elasticity Estimation
with Medical Tools}

\author*[1]{S. Grube}
\author[2]{M. Neidhardt} 
\author[2]{A.-K. Hermann}
\author[2]{J. Sprenger} 
\author[3]{K. Abdolazizi} 
\author[2]{S. Latus}
\author[3]{C. J. Cyron} 
\author[2]{A. Schlaefer} 
\runningauthor{S.~Grube et al.}

\affil[1]{\protect\raggedright 
  Hamburg University of Technology, Institute of Medical Technology and Intelligent Systems, Hamburg, Germany, \newline e-mail: sarah.grube@tuhh.de}
\affil[2]{\protect\raggedright
  Institute of Medical Technology and Intelligent Systems, Hamburg University of Technology, Hamburg, Germany}
  \affil[3]{\protect\raggedright
 Institute for Continuum and Material Mechanics, Hamburg University of Technology, Hamburg, Germany}

\abstract{
    Soft tissue elasticity is directly related to different stages of diseases and can be used for tissue identification during minimally invasive procedures. By palpating a tissue with a robot in a minimally invasive fashion force-displacement curves can be acquired.
    However, force-displacement curves strongly depend on the tool geometry which is often complex in the case of medical tools. Hence, a tool calibration procedure is desired to directly map force-displacement curves to the corresponding tissue elasticity.
    We present an experimental setup for calibrating medical tools with a robot. First, we propose to estimate the elasticity of gelatin phantoms by spherical indentation with a state-of-the-art contact model. We estimate force-displacement curves for different gelatin elasticities and temperatures. Our experiments demonstrate that gelatin elasticity is highly dependent on temperature, which can lead to an elasticity offset if not considered. Second, we propose to use a more complex material model, e.g., a neural network, that can be trained with the determined elasticities. Considering the temperature of the gelatin sample we can represent different elasticities per phantom and thereby increase our training data. 
    We report elasticity values ranging from \SI{10}{} to \SI{40}{kPa} for a \SI{10}{\percent} gelatin phantom, depending on temperature.     
    
 }


\maketitle

\section{Introduction} 
\begin{figure}
    \centering
	\includegraphics[width=0.9\columnwidth]{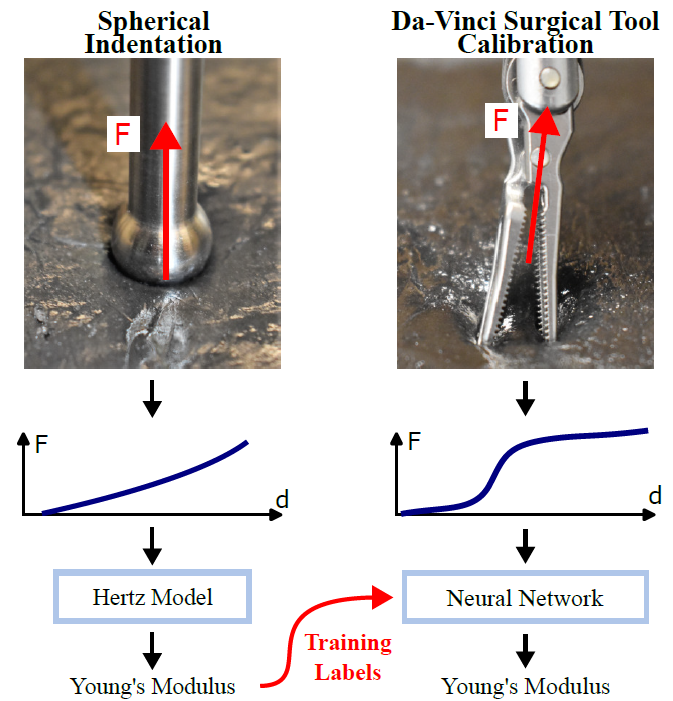}
	\caption{\textbf{Procedure for Calibrating a Surgical Tool.} We propose to map the force-displacement curve to elasticity values with a neural network. As training targets we use the elasticities estimated from indentation experiments with a known geometry (steel ball).}
	\label{fig:Story}
         \vspace{-0.1cm}
\end{figure}

Estimating soft tissue elasticity is helpful during robotic surgery for navigation and localization of surgical tools as well as disease staging~\cite{janmey2011mechanisms}. Commonly, elasticity is estimated by the surgeon by palpating the soft tissue by hand. However, during minimally invasive robotic surgery this is not feasible. Still, the indentation depth and the force between the surgical tool and tissue can be recorded by minimal invasive force sensors integrated inside the surgical tool \cite{fontanelli2020external} or through image-based force estimation \cite{gessert2018force, neidhardt2023optical}. Subsequently, a material model can be applied to estimate the elasticity by approximating the slope of the force-displacement curve. However, tissue deformation and the force recorded are directly related to the tool geometry and might abruptly change during indentation owing to the change of surface in contact with the complex tool geometry as shown in Figure~\ref{fig:Story}. Hence, surgical tools need to be calibrated in advance for accurate elasticity estimation and a model needs to be established that maps the complex force-displacement curve to quantitative elasticity values.

In this work, we present an experimental approach to calibrate complex surgical tools to estimate soft tissue elasticity. Our setup includes a robot with a positioning accuracy of~<~\SI{20}{\micro\meter} which allows us to efficiently perform indentations. We use gelatin phantoms to mimic soft tissue properties, which are easy and inexpensive to manufacture. We estimate the elasticity of our phantoms by performing indentations with a steel ball. By knowing the exact geometry of the spherical indenter we can apply a simple indentation model to estimate the elasticity based on the force curve~\cite{czerner2015determination}, as shown in Figure~\ref{fig:Story}. Thereby, we can estimate the elasticity of our phantoms with high accuracy. Further, we systematically evaluate the influence of different phantom temperatures on elasticity estimates. We hypothesize, that Young's Modulus is dependent on the phantom temperature~\cite{dunmire2013characterizing} and allows us to represent different elasticity values with a single gelatin concentration. Following this, we can efficiently increase the size of our calibration data set by performing multiple indentations on a single phantom that adapts to the room temperature over time. Finally, with our experimental setup surgical tools with complex geometries can be calibrated on known elasticity values of gelatin phantoms as shown in Figure~\ref{fig:Story}, right. Using the estimates from spherical indentation as ground truth, more complex models, e.g., neural networks, which represent the specific surgical tool characteristics can be trained to predict the Young's Modulus from force-displacement curves during minimal invasive surgery. Thereby, acquiring large indentation data sets with a robot, and knowing the ground truth elasticity value, would be beneficial for training neural networks.

In this work, we want to present and evaluate the accuracy of our experimental setup and demonstrate the influence of temperature on gelatin elasticity.

\section{Material and Methods}
\begin{figure}[tb]
	\centering
	\includegraphics[width=0.9\columnwidth]{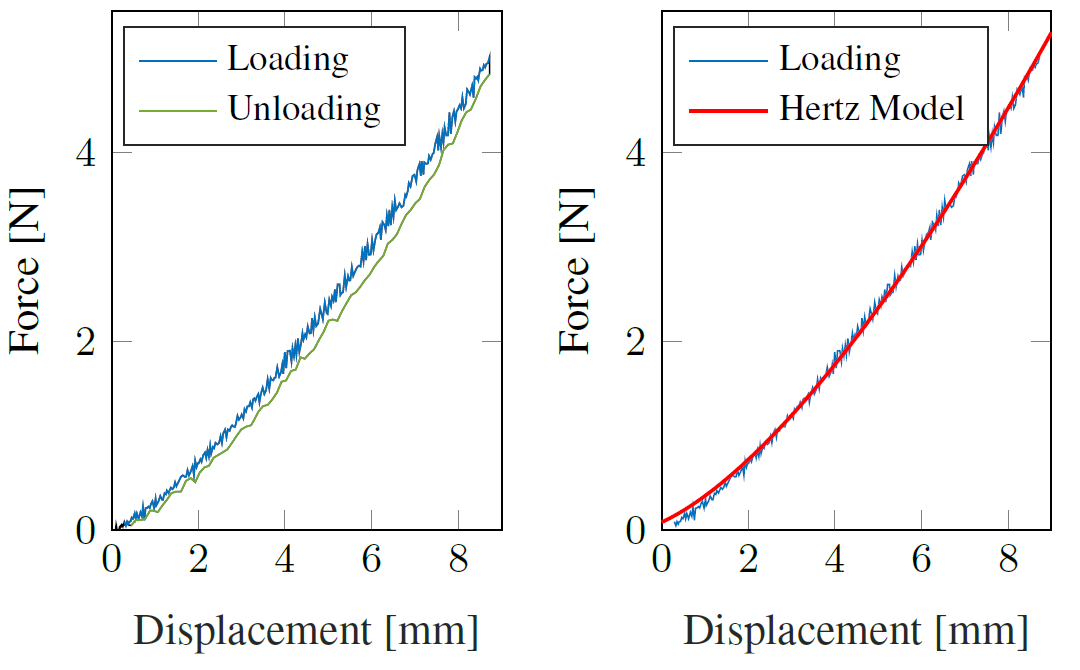}	
 \vspace{-0.1cm}
	\caption{
		\textbf{Loading Curves:} Example of a force-displacement curve for an indentation experiment with a loading and unloading part (Left); the curve fitted based on the Hertz model for estimating the Young's Modulus~$E$ (Right).}	
	\label{fig:MethodErEstimation}
        \vspace{-0.1cm}
\end{figure}

\begin{figure}[tb]
		\centering
            \includegraphics[width=0.9\columnwidth]{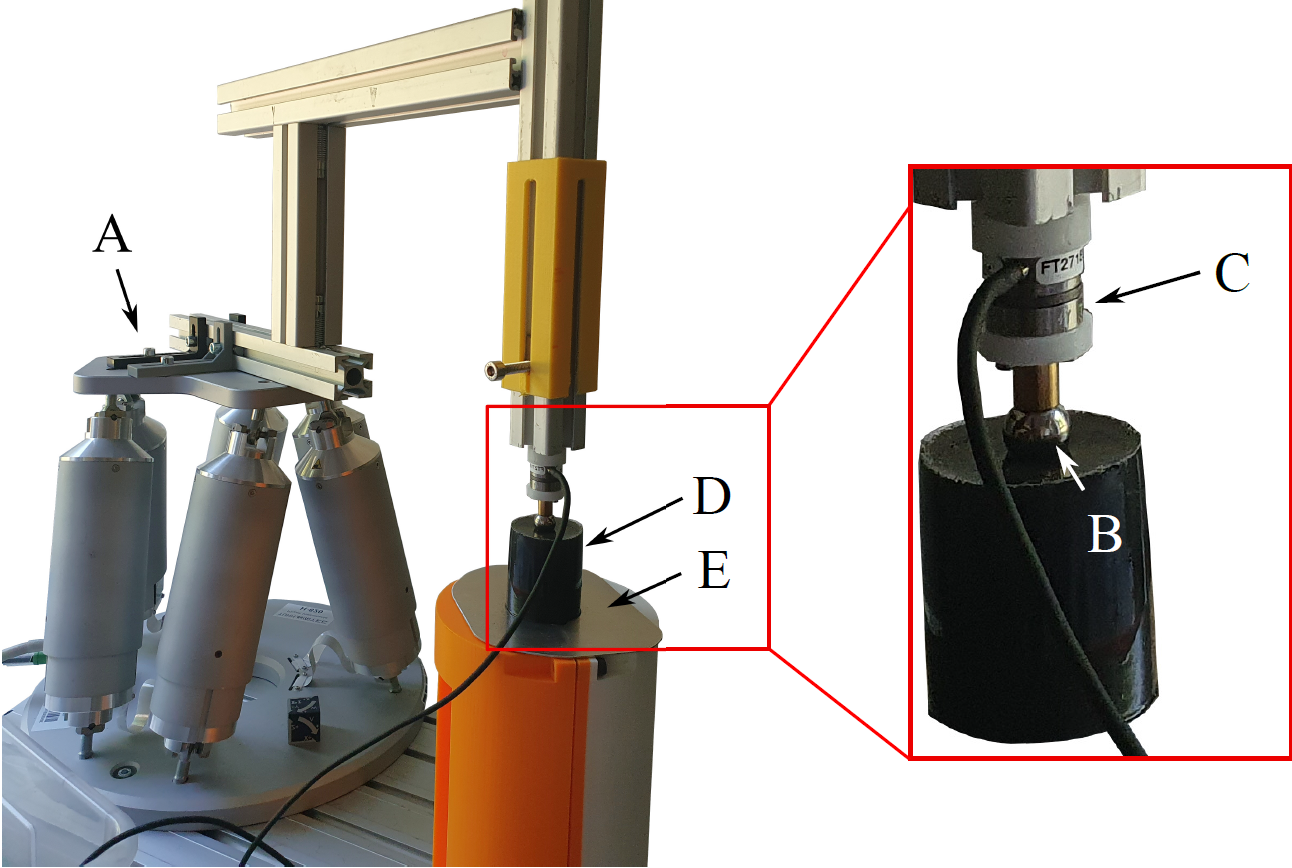}		
		\caption{\textbf{Experimental Setup:} Indentations are performed with a robot~(A). The spherical indenter (B) is attached to the robot via a rigid beam and the phantom (D) is placed on a phantom table (E). Between the indenter and beam a force sensor (C) is mounted.}	
		\label{fig:ExpSetup}
        \vspace{-0.1cm}
\end{figure}

We systematically investigate the influence of gelatin concentration and phantom temperature on the mechanical properties of gelatin phantoms during indentation experiments.  
Each indentation experiment consists of a loading part~(insertion of indenter) and an unloading part~(removal of indenter). The loading and unloading part of a force curve during an indentation is shown in Figure~\ref{fig:MethodErEstimation}. 
Our experimental setup is depicted in Figure~\ref{fig:ExpSetup}.
To palpate tissue phantoms we use a spherical indenter with a radius~$R_\text{ind}$ of \SI{6}{\milli\meter}. The indenter is attached to a force sensor~(Nano17, ATI, USA) to record the forces during indentation. We use a robot~(Hexapod H-820, PI, DE) to automatically perform indentations with an axial repeatability precision of \SI{20}{\micro\meter}. 
The indenter is pressed into the phantom until a maximum insertion distance of \SI{10}{\milli\meter} is reached or a maximum force of \SI{5}{\newton}. In our experiments we vary gelatin concentration and phantom temperature.
We repeat each indentation experiment ten times. For each repetition, a new phantom sample is used to avoid inaccuracies due to changes in the mechanical properties caused by previous indentations. Furthermore, all of the ten phantoms are made from the same batch of gelatin to ensure the same gelatin properties within the repetition experiments. 
We store the gelatin phantoms for \SI{24}{h} in the refrigerator for hardening at a temperature of \SI{7}{\celsius}. To investigate the influence of phantom temperature, we perform our indentation experiments directly after removal from the refrigerator (\SI{7}{\celsius}) and compare the results with indentations performed at gelatin phantoms that have already been exposed to the ambient temperature for a longer time and thus are warmer~(\SI{15}{\celsius} and \SI{22}{\celsius}). 
In addition to the parameter study, we also perform a long-term study in which we repeat indentations every \SI{10} minutes on a \SI{10}{\%} gelatin phantom for \SI{2}{h}. We repeat the long-term study with five phantoms at different ambient temperatures to show dependence on temperature. During each indentation we measure the inner phantom temperature. \textcolor{black}{Note, since our goal is only to obtain different force-displacement curves, we do not consider the temperature gradient between phantom surface and the center. This may affect variance, but is assumed to be likely the same for similar shape.}

\subsection{Data Evaluation}
To evaluate mechanical properties of soft tissue the Young's Modulus $E$  is estimated based on the Hertz-Model~\cite{czerner2015determination} which assumes a homogeneous, isotropic and linear elastic material.
The Hertz model fits a curve 
\begin{equation}
	F = \frac{4}{3} \frac{E}{1-\nu^2} R_\text{ind}^{\frac{1}{2}} d^{\frac{3}{2}}
\end{equation}
to the loading force-displacement curve with an indentation depth~$d$ and Poisson ratio~$\nu=0.5$. The formula assumes a spherical indenter with a radius~$R_\text{ind}$. In Figure~\ref{fig:MethodErEstimation}, right, the curve fitting of the loading curve is shown. 

\section{Results and Discussion}
\begin{figure}[tb]
	\centering
	\includegraphics[width=0.9\columnwidth]{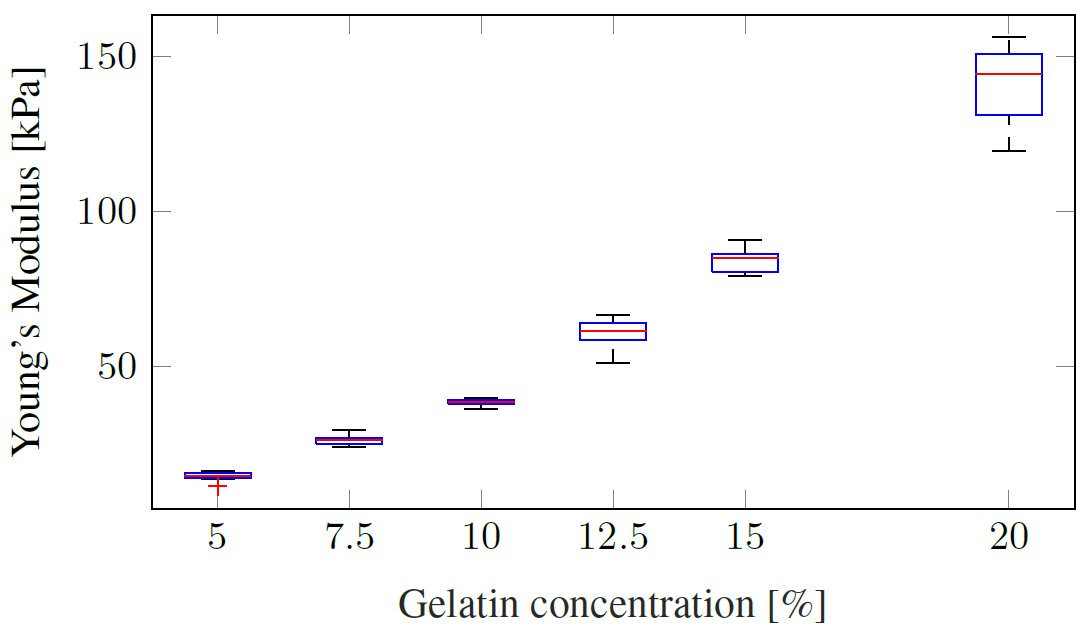}
        \vspace{-0.2cm}
	\caption{\textbf{Spherical Indentation Experiments:} boxplots showing the results of the performed spherical indentation experiments to study the influence of gelatin concentration on Young's Modulus.}	
	\label{fig:InfluenceGelParameters}
        \vspace{-0.1cm}
\end{figure}

\begin{figure}[tb]
	\centering
	\includegraphics[width=0.9\columnwidth]{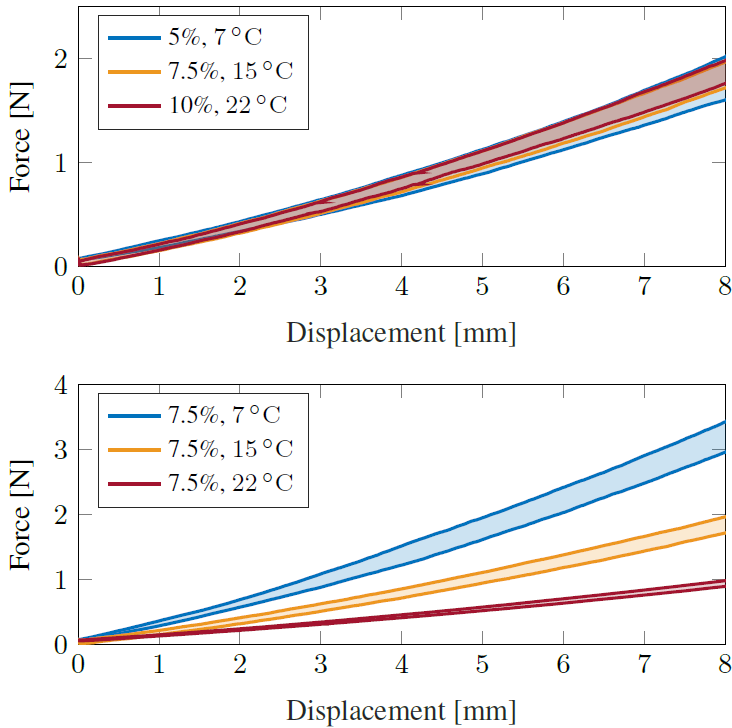}
 \vspace{-0.2cm}
	\caption{\textbf{Force-Displacement Curves:} Loading curves for different gelatin concentrations and temperatures. (Top) Depicted are \SI{5}{\%}, \SI{7.5}{\%} and \SI{15}{\%} gelatin phantoms with \SI{7}{\celsius}, \SI{15}{\celsius} and \SI{22}{\celsius}, respectively. (Bottom) The loading curves for 7.5\% gelatin phantoms at \SI{7}{\celsius}, \SI{15}{\celsius} and \SI{22}{\celsius}.}	
	\label{fig:diffGelTempInflu}
 \vspace{-0.2cm}
\end{figure}

\begin{figure}[tb]
    \centering
\includegraphics[width=0.9\columnwidth]{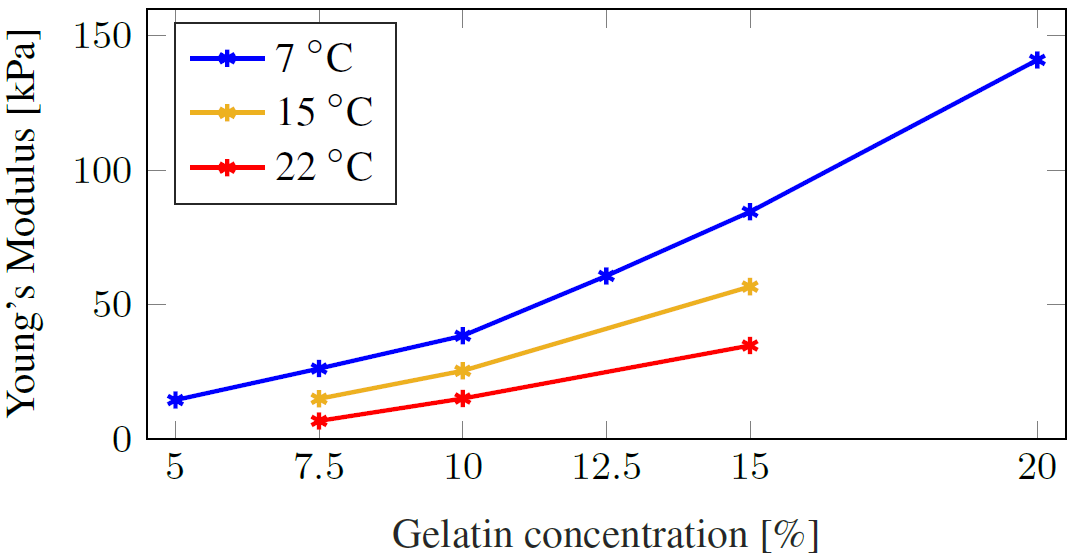}
\vspace{-0.2cm}
	\caption{\textbf{Young's Modulus vs. Temperature:} Young's Modulus estimates for gelatin concentrations and phantom temperatures of \SI{7}{\celsius}, \SI{15}{\celsius} and \SI{22}{\celsius}.}	
	\label{fig:InfluenceParametersTemp}
  \vspace{-0.3cm}
\end{figure}

The results of the parameter study are visualized in Figure~\ref{fig:InfluenceGelParameters} and Figure~\ref{fig:InfluenceParametersTemp}. We test for significant differences in the Young's Modulus using the Wilcoxon signed-rank test with a significance level of \SI{5}{\%}.
An increase in gelatin concentration leads to an increase in the Young's Modulus (Figure \ref{fig:InfluenceGelParameters}). The gelatin concentrations we studied have a significantly different Young's Modulus and lie in the range of soft tissue ~\cite{neidhardt2022ultrasound}.
Increasing the phantom temperature leads to a decrease in the Young's Modulus~(Figure~\ref{fig:InfluenceParametersTemp}).
Our results show that phantoms with different gelatin concentrations can have the same Young's Modulus depending on their temperature~(Figure \ref{fig:InfluenceParametersTemp}). 
The Young's Modulus of a \SI{7.5}{\percent} gelatin phantom at \SI{7}{\celsius} does not significantly differ from a \SI{10}{\percent} gelatin phantom at \SI{15}{\celsius}.
Additionally, no significant differences in the Young's Modulus can be observed, when comparing the overlapping loading curves from \SI{5}{\percent}, \SI{7.5}{\percent} and \SI{10}{\percent} gelatin phantom with \SI{7}{\celsius}, \SI{15}{\celsius} and \SI{22}{\celsius} phantom temperature, respectively~(Figure~\ref{fig:diffGelTempInflu}, top).
For comparison, Figure~\ref{fig:diffGelTempInflu}, bottom, shows the standard deviation of the loading curves of \SI{7.5}{\percent} gelatin phantoms at different temperatures. In contrast to Figure~\ref{fig:diffGelTempInflu}, top, the regions of the force-displacement curves differ from each other, depending on the phantom temperature which leads to a significant difference in the Young's Modulus.

\begin{figure}[tb]
	\centering
\includegraphics[width=0.9\columnwidth]{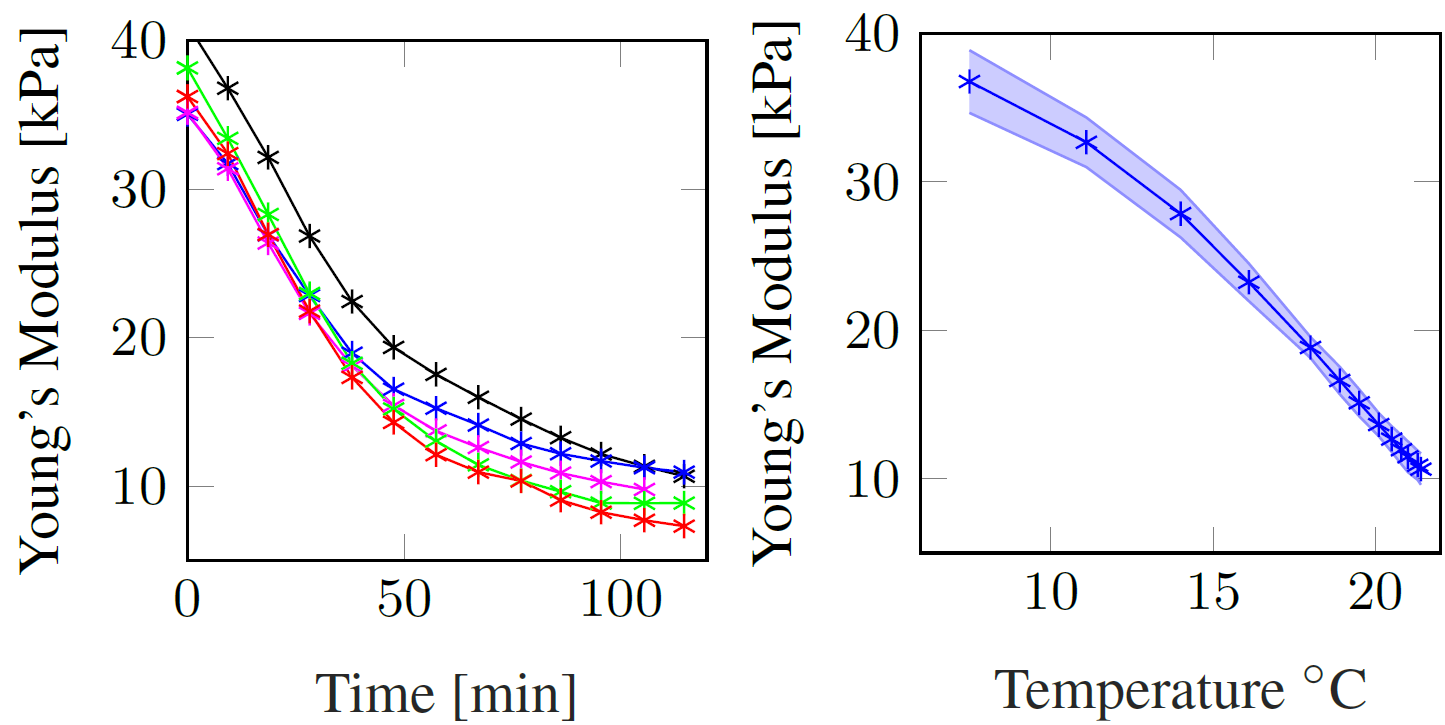}
 \vspace{-0.2cm}
	\caption{\textbf{Young's Modulus vs. Temperature for 10\% Gelatin:} (Left) Young's Modulus estimates at different time points after removing from cooling. Each color indicates an individual phantom. (Right) Dependency between the Young's Modulus and temperature. The shaded surface indicates the standard deviation. }	
	\label{fig:timeTempDepend}
 \vspace{-0.3cm}
\end{figure}

When performing 12 indentation experiments every 10 minutes on a \SI{5}{\%} gelatin phantom the Young's Moduli are ranging from \SI{9.61}{kPa} to \SI{43.13}{kPa} (Figure~\ref{fig:timeTempDepend}). The relationships between the Young's modulus and the gelatin temperature or time outside the refrigerator are shown in Fig.~\ref{fig:timeTempDepend}.
Evaluating the Young's Modulus over all measurements by taking only the gelatin concentration into account, a mean standard deviation of \SI{9.26}{kPa} results with differences up to \SI{33.79}{kPa}. 
When considering the time outside the refrigerator we already can reduce the mean standard deviation to \SI{2.01}{kPa}. However, when taking the phantom temperature and thus also the ambient temperature into account the mean standard deviation is reduced to \SI{1.14}{kPa}~(Figure~\ref{fig:timeTempDepend}, right).  \textcolor{black}{It is noticeable, that larger standard deviations occur at lower temperatures. This might be due to a larger delta to the ambient temperature.}
Based on the long-term study, we establish an equation to estimate the phantom stiffness of a \SI{10}{\percent} gelatin concentration as a function of phantom temperature $T_\text{ph}$. Quadratic fitting yields in 
\begin{equation}
	E=-0.06 T_\text{Ph}^2-0.13 T_\text{Ph}+41.42
\end{equation}
with a coefficient of determination of $R^2=0.9992$ considering a temperature range of \SI{7.5}{\celsius} up to \SI{21.4}{\celsius}. 

As motivated in Figure~\ref{fig:Story}, our calibration approach can be used to train a neural network that predicts elasticity estimates from force-displacement curves generated by complex medical tools. We demonstrate that gelatin concentration as well as the gelatin phantom temperature are systematically related to the Young's Modulus. Hence, it can be concluded that a change in phantom temperature allows us to easily model different stiffness values from a single phantom. 
This not only leads to a substantial reduction of time needed to generate gelatin phantoms. It also allows to map more stiffness values by continuously performing indentation experiments while the phantom adjusts to ambient temperature. Furthermore, we have shown that we can reduce inaccuracies in the ground truth values by taking the phantom temperature into account. Thus, a huge and precise calibration data set can be acquired for training of neural networks.

\section{Conclusion}
We present an experimental setup for acquiring calibration data to estimate elasticity from palpation with a surgical tool. Our results show that we can model different tissue stiffnesses by changing either the phantom temperature or the gelatin concentration. Phantoms with gelatin concentrations from \SI{5}{\percent} up to \SI{20}{\percent} can be used to model Young's Moduli which lie in the range of \SI{7}{kPa} up to \SI{150}{kPa}. 
We have shown that it is important to measure the phantom temperature during data acquisition as it has a significant impact on the Young's Modulus. Ignoring the phantom temperature leads to less accurate ground truth values. 
Furthermore, we have shown that the temperature-dependent stiffness of gelatin phantoms has not only disadvantages since there is a systematic relationship to phantom stiffness. 
Considering the phantom temperature, we can perform indentation experiments on phantoms with different stiffness values in a simpler and more effective way than if we map different stiffnesses by changing the gelatin concentration.
By palpating a single phantom as it adjust to ambient temperature, force-displacement curves of varying stiffness can be recorded due to the temperature-dependent material properties.
Our calibration approach could be used to train a neural network for predicting tissue stiffness based on tool-tissue interaction forces with a large and variable data set.
\vspace{-0.5cm}

\subsection*{Author Statement}
Research funding: This work was partially funded by TUHH $\text{i}^3\text{m}^4$ initiative and Interdisciplinary Competence Center for Interface Research (ICCIR). 
Conflict of interest: Authors state no conflict of interest. Informed consent: Informed consent has been obtained from all individuals included in this study. Ethical approval: No ethical approval was necessary for this research.

\vspace{-0.5cm}

\bibliographystyle{elsevier3}
\bibliography{ref}

\end{document}